\UseRawInputEncoding
\documentclass[reprint,
superscriptaddress,
longbibliography,
amsmath,amssymb,aps,pra]{revtex4-2}
\usepackage{cuted}
\usepackage[dvipsnames]{xcolor}
\usepackage{graphicx}
\usepackage{dcolumn}
\usepackage{bm}
\usepackage[mathlines]{lineno}
\usepackage{hyperref}
\hypersetup{colorlinks=true,linkcolor=blue,citecolor=blue,urlcolor=blue}
\usepackage{graphicx}
\usepackage{lipsum}
\begin{document}

\title{Experimental demonstration of enhanced violations of Leggett-Garg \\inequalities in a $\mathcal{PT}$-symmetric trapped-ion qubit}
\author{Pengfei Lu}
\thanks{These authors contributed equally to this work.}
\affiliation{School of Physics and Astronomy, Sun Yat-Sen University, Zhuhai, 519082, China}

\author{Xinxin Rao}
\thanks{These authors contributed equally to this work.}
\affiliation{School of Physics and Astronomy, Sun Yat-Sen University, Zhuhai, 519082, China}

\author{Teng Liu}
\affiliation{School of Physics and Astronomy, Sun Yat-Sen University, Zhuhai, 519082, China}

\author{Yang Liu}
\affiliation{School of Physics and Astronomy, Sun Yat-Sen University, Zhuhai, 519082, China}
\affiliation{Shenzhen Research Institute of Sun Yat-Sen University, Nanshan Shenzhen 518087, China}

\author{Ji Bian}
\affiliation{School of Physics and Astronomy, Sun Yat-Sen University, Zhuhai, 519082, China}

\author{Feng Zhu}
\affiliation{School of Physics and Astronomy, Sun Yat-Sen University, Zhuhai, 519082, China}
\affiliation{Shenzhen Research Institute of Sun Yat-Sen University, Nanshan Shenzhen 518087, China}
\author{Le Luo}
\email{luole5@mail.sysu.edu.cn}
\affiliation{School of Physics and Astronomy, Sun Yat-Sen University, Zhuhai, 519082, China}
\affiliation{Shenzhen Research Institute of Sun Yat-Sen University, Nanshan Shenzhen 518087, China}
\affiliation{State Key Laboratory of Optoelectronic Materials and Technologies, Sun Yat-Sen University, Guangzhou 510275, China}
\affiliation{International Quantum Academy, Shenzhen, 518048, China}

\begin{abstract}
The Leggett-Garg inequality (LGI) places a bound for the distinction between quantum systems and classical systems. Despite that the tests of temporal quantum correlations on LGIs have been studied in Hermitian realm, there are still unknowns for LGIs in non-Hermitian conditions due to the interplay between dissipation and coherence. For example, a theoretical hypothesis to be experimentally validated, suggests that within non-Hermitian systems, the non-unitary evolution of the system dynamics allows the boundaries of the LGIs to surpass the constraints imposed by traditional quantum mechanics. Here, we demonstrate the experimental violation of LGIs in a parity-time ($\mathcal{PT}$)-symmetric trapped-ion qubit system by measuring the temporal correlation of the evolving states at different times. We find that the upper bounds of the three-time parameter $K_3$ and the four-time parameter $K_4$ show enhanced violations with the increasing dissipation, and can reach the upper limit by infinitely approaching exceptional point. We also observe the distinct behavior of the lower bounds for $K_3$ and $K_4$. While the lower bound for $K_3$ remains constant, the case for $K_4$ shows an upward trend with increasing dissipation. These results reveal a pronounced dependence of  the system's temporal quantum correlations on its dissipation to the environment. This opens up a potential pathway for harnessing dissipation to modulate quantum correlations and entanglement.
\end{abstract}

\maketitle

\section{introduction}
Quantum state superposition~\cite{schrodinger1926quantisierung} is a fundamental concept in quantum mechanics and plays a crucial role in various applications, including quantum computing~\cite{steane1998quantum}, quantum simulation~\cite{georgescu2014quantum}, and quantum communication~\cite{gisin2007quantum}. To illustrate the counterintuitive property of the superposition, in 1935, Schr{\"o}dinger proposed the famous ``Schr{\"o}dinger's cat" paradox~\cite{schrodinger1935gegenwartige}. In order to investigate the validity of this paradox in experiments, Leggett and Garg proposed the Leggett-Garg inequality (LGI) in 1985~\cite{leggett1985quantum}. This inequality is based on the assumption of macroscopic realism (MR) and non-invasive measurement (NIM) and aims to study the temporal correlation of the system evolution.

In quantum mechanics, the principle of quantum superposition and the collapse of quantum states violate the MR and the NIM assumption, respectively. As a result, the violation of the LGI can occur in quantum systems. The experimental tests of the LGIs have primarily focused on closed quantum systems, such as superconducting circuits~\cite{palacios2010experimental}, single-photon~\cite{zhou2015experimental,dressel2011experimental,wang2017enhanced}, nuclear magnetic resonance system~\cite{athalye2011investigation,katiyar2013violation}, trapped-ion~\cite{zhan2023experimental}, and NV center~\cite{waldherr2011violation,tusun2022experimental} etc. However, the interplay between a quantum system and its external environment gives rise to the complex evolutions, such as the chiral, nonreciprocal state transfer~\cite{liu2021dynamically,abbasi2022topological} and the inhomogeneity of state evolution speed~\cite{lu2022realizing,varma2022essential}. This uneven state evolution may result in enhanced violations of the upper bound on the LGIs~\cite{karthik2021leggett,varma2021temporal,varma2022essential}. Thanks to the improvement of quantum control technology in the atom level, engineering an open quantum system with controllable dissipation can be realized~\cite{barreiro2011open,kienzler2015quantum,harrington2022engineered,ding2021experimental,bian2022quantum}. Recently, the experimental investigations of the LGI in open quantum systems have been studied in trapped-ion systems~\cite{wu2023maximizing,quinn2023observing}. They used the parameter $K_3$ (LGI order n=3) to account for the enhanced quantum correlation by approaching an exceptional point. Different with these results, we concentrate on the variations in the bounds of  $K_3$ and the higher-order $K_4$ (LGI order n=4) in relation to the dissipation, and show how the upper and lower bounds (especially $K_4$) are influenced by the target state under certain observable operator.

In this work, we systematically investigate the three and four order LGIs ($K_3$ and $K_4$), which act as the temporal counterparts of the well-known Bell~\cite{bell1964einstein} and CHSH~\cite{clauser1969proposed} inequalities, by utilizing the trapped-ion system with a $\mathcal{PT}$-symmetric Hamiltonian~\cite{bender1998real,li2019observation}. We demonstrate the enhanced violations in the upper bound of  these LGIs with the increasing dissipation. But these LGIs exhibit a different behavior for the lower bound. Specifically, for $K_3$, it remains constant, while for $K_4$, the lower bound is shifted depending on the dissipation. The distinct nature in the odd-even LGIs suggests that the different order of LGIs could be used to explore various violations of the macrorealism in open quantum systems. For example, the reducible higher-order LGIs are useful in addressing the clumsiness loophole~\cite{wilde2012addressing} and the irreducible one from the cut polytope can offer a nontrivial way to test the macrorealism~\cite{avis2010leggett}.

\section{Experimental implementation}
\begin{figure}
	\includegraphics[width=1.0\linewidth]{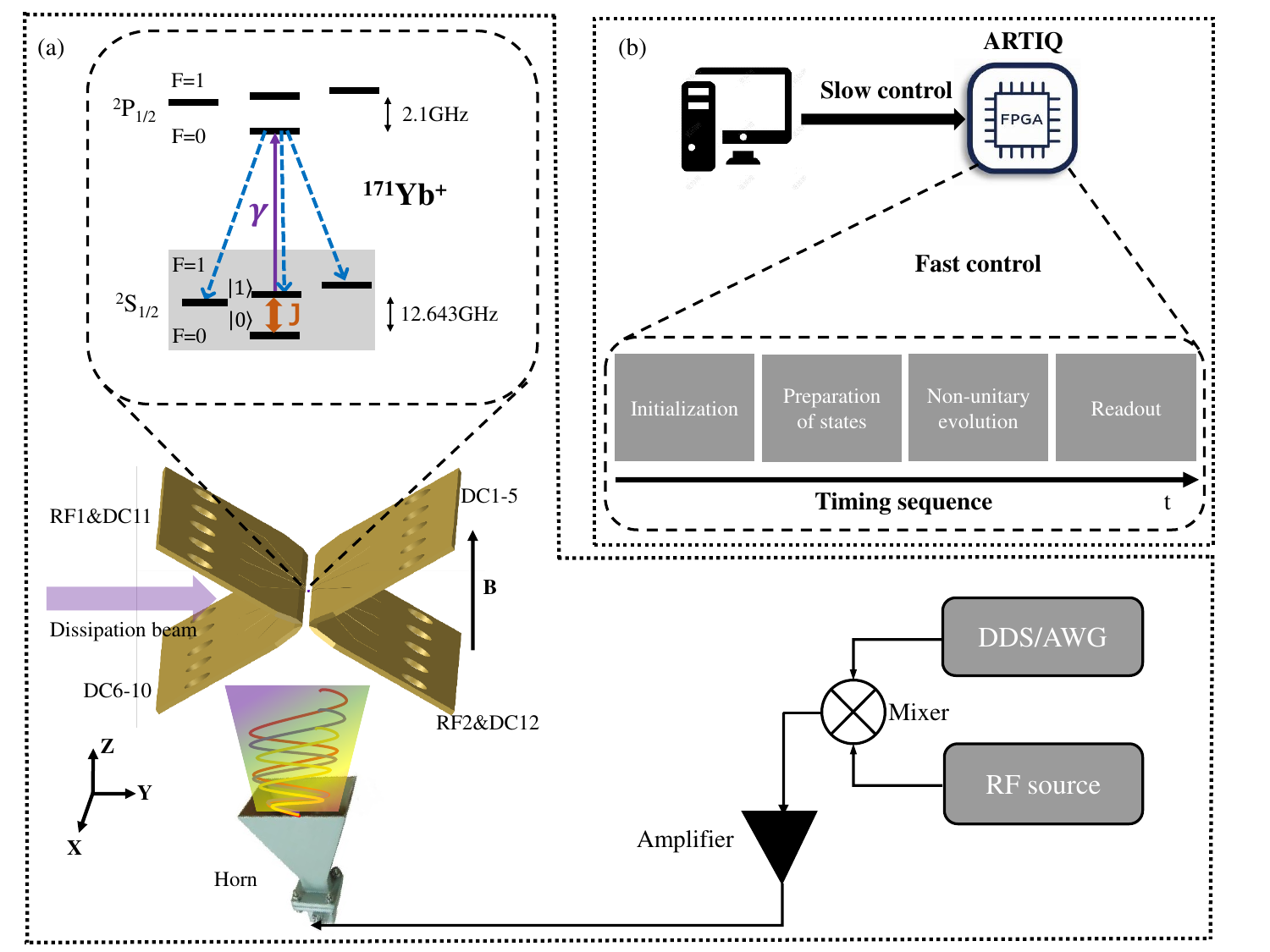}
	\caption{Experiment setup for testing LGIs. (a) The ion trap used in this experiment consists of four gold-plated ceramic blade electrodes, which confine the ion near the center of the trap. The configuration of the microwave and dissipation beam setup is illustrated. The generation of the 12.643 GHz microwave signal by mixing a standard RF source with either a direct digital synthesizer (DDS) or an arbitrary waveform generator (AWG). The resulting signal is then amplified to drive the transition between the states $|0\rangle$ and $|1\rangle$ by microwave horn. The energy level diagram of $^{171}\text{Yb}^+$ ion is depicted within the dashed box. A 369 nm laser contains only $\pi$ polarization component (purple line) is used to excite the ion from the ground state $|1\rangle$ to the $F=0$ state of the $ ^{2}P_{1/2}$ level, and then spontaneously radiates (blue dashed line) and returns to the $F=1$ state of the $^{2}S_{1/2}$ level . The gray shaded region within the $ ^{2}S_{1/2}$ manifold can be approximated as a dissipative two-level system. (b) Experimental timing sequences. Initially, the ion is optically pumped to the state $|0\rangle$. Next, we prepare the target state and allow for non-unitary evolution for a certain time. Finally, we employ the standard fluorescence counting rate threshold method to readout experimental results.}
	\label{experiment_setup}
\end{figure}
Here, we consider the non-unitary evolution of the two level system driven by the effective non-Hermitian Hamiltonian~\cite{lu2022realizing,ding2021experimental}, which can be described as ($\hbar$=1),
\begin{equation}
	H_{eff}=\begin{pmatrix}
		0 & J\\
		J &-2i\gamma
	\end{pmatrix},
	\label{Heff}
\end{equation}
where $J$ is the coupling strength and $\gamma$ is the dissipation rate. Eq.~(\ref{Heff}) can be mapped to the $\mathcal{PT}$-symmetric Hamiltonian by adding an identity matrix, $H_{\mathcal{PT}}=H_{eff}+i\gamma \bm{I}$. The eigenvalues of $H_{\mathcal{PT}}$ are $\pm\sqrt{J^2-\gamma^2}$, where the exceptional point locates at $J=\gamma$.

In order to realize Eq.~(\ref{Heff}), our experiments are implemented on a trapped-ion platform. A single $^{171}$Yb$^+$ ion is confined in a linear blade Paul trap. The schematics of the trap and energy levels of the ion are depicted in Fig.~\ref{experiment_setup}. The ion can be initialized to either $|0\rangle=|F = 0, m_F = 0\rangle$ or
$|1\rangle=|F = 1, m_F = 0\rangle$ in $^{2}S_{1/2}$ in the ground state hyperfine manifold. A microwave with the frequency of 12.643 GHz is used to couple the two spin states. A 369.5 nm laser beam with $\pi$ polarization and adjustable intensity is employed as a dissipation beam to excite the ion from $|1\rangle$  to  $^{2}P_{1/2}$ electronic excited state, which leads to spontaneous decay to three magnetic levels $|F = 1, m_{F} = 0, \pm 1\rangle$ in $^{2}S_{1/2}$ ground state with equal probability. The decay to $|F = 1, m_{F} = \pm 1\rangle$ can be considered as equivalent loss of the qubit, resulting in a non-unitary evolution~\cite{ding2021experimental,bian2022quantum}. The detailed experimental setup can be found in Appendix~\ref{appA}.
		
\section{METHODS AND RESULTS}

\begin{figure}
	\includegraphics[width=0.9\linewidth]{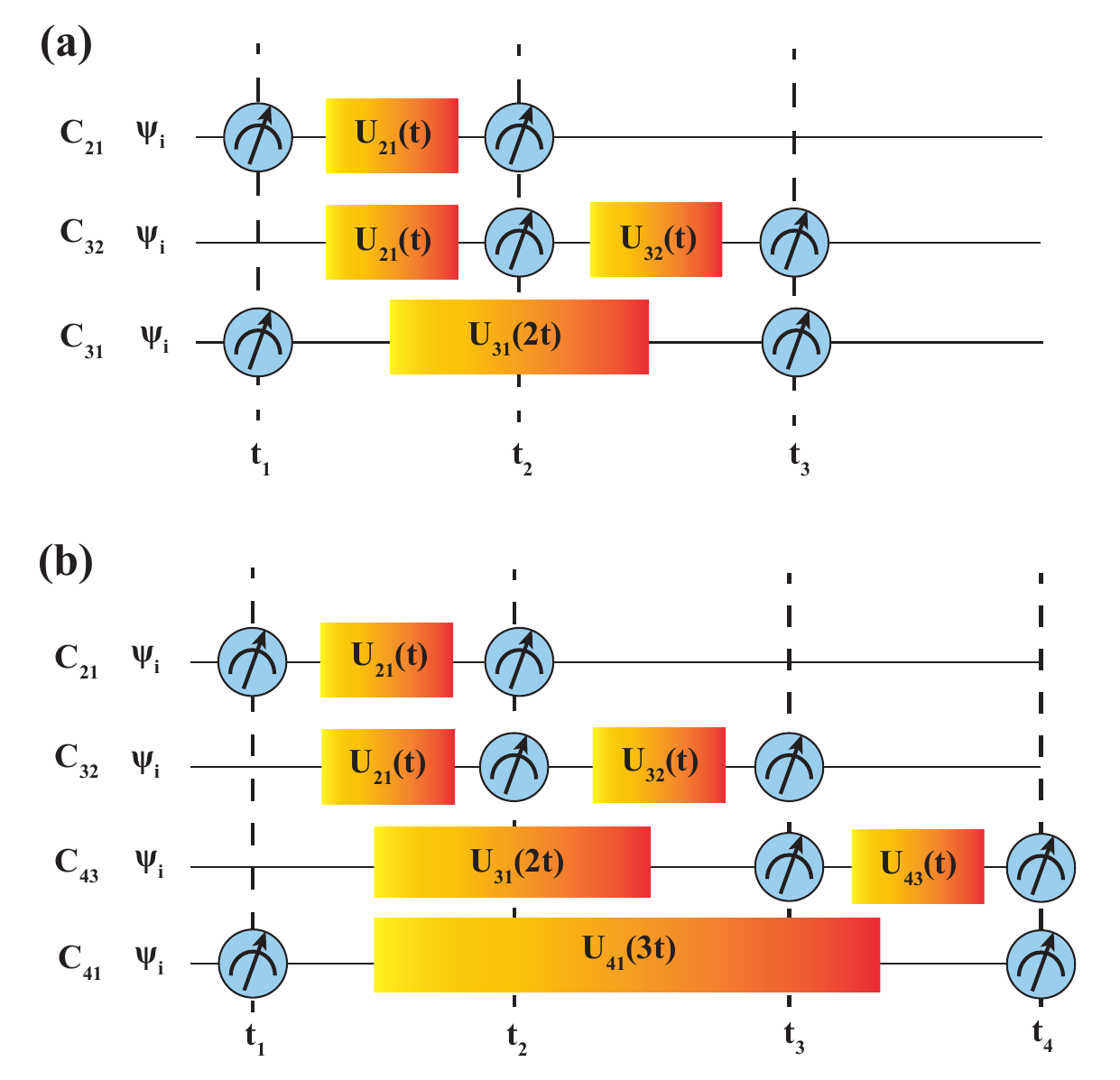}
	\caption{Schemes for the LGI test of the three-time ($K_3$) and four-time ($K_4$) scenario. (a) The experiments of three two-time temporal correlations ($C_{21}$, $C_{32}$, and $C_{31}$) need to be performed to obtain the parameter $K_3$. (b) The experiments of four two-time temporal correlations ($C_{21}$, $C_{32}$, $C_{43}$, and $C_{41}$) need to be performed to obtain the parameter $K_4$. $|\psi_i\rangle$ represents the state $|1\rangle$.}
	\label{K34_scheme}
\end{figure}
Unlike the Bell inequality, which is used to test correlations between spatially separated systems~\cite{franson1989bell}, the LGI is used to test correlations within a single system at different moments. The most frequently encountered LGI has the following general form~\cite{emary2013leggett},
\begin{equation}
K_{n}=C_{21}+C_{32}+C_{43}+\dots+C_{n(n-1)}-C_{n1},
\label{equ1}
\end{equation}
where $C_{n(n-1)}$ is the correlation function of observables between the times $ t_{n} $ and $ t_{n-1}$. Typical three and four-time LGIs denote as $K_ 3=C_{21}+C_{32}-C_{31}$ and $K_{4}=C_{21}+C_{32}+C_{43}-C_{41}$ where the value of $K_ 3$ ($K_ 4$) is bound by $ -3\le K_{3}\le 1$ ($ -2\le K_{4}\le 2$). Schemes for the LGI test of the three-time ($K_3$) and four-time ($K_4$) scenario are shown in Fig.~\ref{K34_scheme}(a-b).

In a two-level quantum system, the upper bound of the $ K_3 $ ($ K_4$) can exceed the value of 1 (2) and reach $3/2$ ($2\sqrt{2}$) according to the L{\"u}der state update rule~\cite{luders2006concerning}. However, the upper bound of $K_3 $ ($ K_4$) could exceed $3/2$ ($2\sqrt{2}$) under von Neumann state upstate rule in a multi-level system~\cite{budroni2014temporal}. It is natural to ask whether the violation of LGI can emerge in a two-level system. Recent studies~\cite{karthik2021leggett,varma2021temporal,varma2022essential} have demonstrated that a two-level system undergoing nonunitary dynamics could lead to the enhanced violations of LGI.

\begin{figure}
	\includegraphics[width=0.9\linewidth]{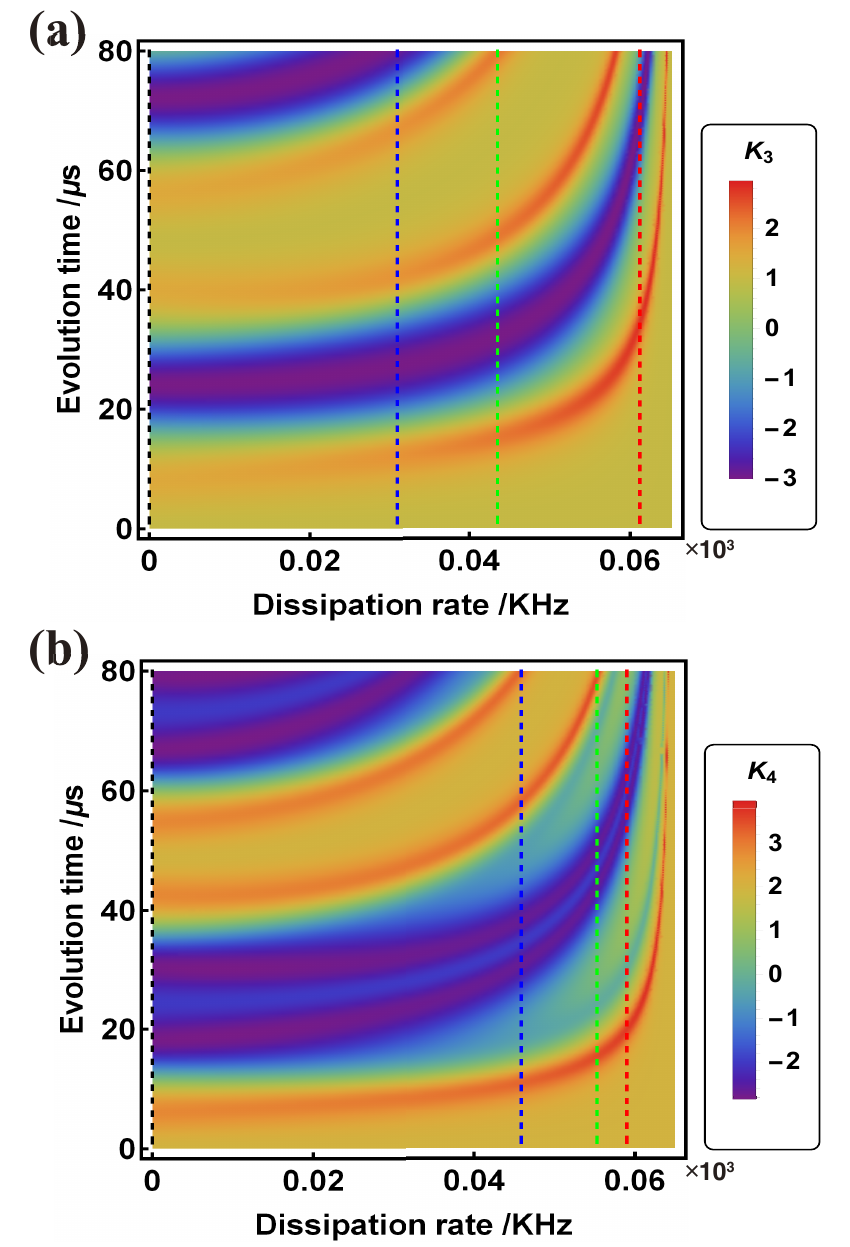}
	\caption{Theoretical plots of the three-time $K_3$ (a) and four-time $K_4$ (b) as a function of $\gamma$ and $\tau$. The black, blue, green, and red dashed lines in (a) represent $\gamma=2\pi\times$0 KHz, $2\pi\times$4.9 KHz, $2\pi\times$7 KHz and $2\pi\times$9.7 KHz, respectively. Similarly, the black, blue, green, and red dashed line in (b) represent $\gamma=2\pi\times$0  KHz, $2\pi\times$7.3 KHz, $2\pi\times$8.9 KHz and $2\pi\times$9.4 KHz, respectively. The coupling strength $J$ in (a) and (b) are 2$\pi\times$10.4 KHz.}
	\label{K34_value}
\end{figure}

We calculate two-time correlations of the observable $Q=\sigma_y$, where the $|+\rangle=\lbrace i/\sqrt{2},1/\sqrt{2}\rbrace^{T} $ and $ |-\rangle= \lbrace -i/\sqrt{2},1/\sqrt{2}\rbrace^{T}$ are the two eigenvectors of the observable operator. The dichotomic outcome $Q_i$ and  $Q_j$ can take a value of 1 or -1 due to the MR assumption. Thus, the correlation function $C_{ij}$ is given by:
\begin{equation}
C_{ij}=\sum_{Q_i,Q_j=\pm1}Q_iQ_jP_{ij}(Q_i,Q_j),
\end{equation}	
where $P_{ij}(Q_i,Q_j)$ denotes the joint probability of observing the measurement outcomes $Q_i$ and $Q_j$ at times $t_i$ and $t_j$, respectively. The four joint probability ($P_{ij}(+,+)$, $P_{ij}(+,-)$, $P_{ij}(-,+)$, $P_{ij}(-,-)$) can be written as,
\begin{equation}
\begin{aligned}
&P_{ij}(\pm,\pm)
\\&=\frac{|\langle \pm| e^{-iH_{eff}t_{i}}|\psi_{t}\rangle |^{2}|\langle \pm| e^{-iH_{eff}t_{ji}}|\pm \rangle|^{2}}{\langle \psi_{t}| e^{iH_{eff}^{\dagger}t_{i}}e^{-iH_{eff}t_{i}}|\psi_{t}\rangle \langle \pm| e^{iH_{eff}^{\dagger}t_{ji}}e^{-iH_{eff}t_{ji}}|\pm \rangle}.
\label{joint}
\end{aligned}	
\end{equation}

The target state $|\psi_t\rangle$ we choose is the $|+\rangle$ and the process from $|\psi_i\rangle$ to $|\psi_t\rangle$ can be regarded as the first measurement (Fig.~\ref{K34_scheme}). The choice of the target state can be obtain from the optimized $K_3$ and $K_4$ values in $\theta$-$\phi$ plane (See Appendix~\ref{appB}). When the equal time spacing $t_4-t_3=t_3-t_2=t_2-t_1=\tau$ is set, the LGI parameters of three and four-time in Hermitian case (the absence of the dissipation rate $\gamma$) is expressed by,	
\begin{equation}
\begin{split}
K_{3}=2\cos(2J\tau)-\cos(4J\tau)
\\
K_{4}=3\cos(2J\tau)-\cos(6J\tau).
\end{split}
\label{k3eq1}
\end{equation}	

While in non-Hermitian case (the presence of the dissipation rate $\gamma$), the LGIs can be replaced by,	
\begin{equation}
\begin{aligned}
K_3&=\frac{\gamma+J\cos(2\tau\chi )}{J+\gamma\cos(2\tau \chi )}-\frac{\gamma+J\cos(4\tau\chi )}{J+\gamma\cos(4\tau\chi )}
\\&+\frac{J\gamma^2+J(J^2+J\gamma-\gamma^2)\cos(2\tau\chi )}{(J-\gamma\cos(2\tau\chi ))(J+\gamma\cos(2\tau\chi ))^2}
\\&-\frac{\gamma\cos(2\tau\chi )(-J^2+J\gamma+\gamma^2+J^2\cos(2\tau\chi ))}{(J-\gamma\cos(2\tau\chi ))(J+\gamma\cos(2\tau\chi ))^2},
\end{aligned}
\label{k3eq}
\end{equation}	
and
\begin{widetext}
\begin{equation}
\begin{aligned}
K_4&=\frac{\gamma+J\cos(2\tau \chi)}{J+\gamma\cos(2\tau \chi)}-\frac{\gamma+J\cos(6\tau\chi)}{J+\gamma\cos(6\tau\chi)}
+\frac{(\gamma +J) \cos ^2(t \chi ) (\gamma +J \cos (2 t \chi ))}{(J+\gamma  \cos (2 t \chi ))^2}-\frac{2 (J-\gamma ) \sin ^2(t \chi ) (J \cos (2 t \chi )-\gamma )}{\gamma ^2-2 J^2+\gamma ^2 \cos (4 t \chi )}
\\&+\frac{2 \left(-\gamma ^3+2 J^3+\gamma  J^2-2 \gamma ^2 J\right) \cos (2 t \chi )+\gamma  \left(2 J (J-\gamma ) \cos (4 t \chi )-J (-2 \gamma +J \cos (8 t \chi )+J)+2 \chi ^2 \cos (6 t \chi )\right)}{4 (J-\gamma  \cos (2 t \chi )) (J+\gamma  \cos (2 t \chi )) (J+\gamma  \cos (4 t \chi ))},
\end{aligned}
\label{k4eq}
\end{equation}	
\end{widetext}
where $\chi=\sqrt{J^2-\gamma^2}$. The color maps of three-time $K_3$ and four-time $K_4$ as a function of $\gamma$ and $\tau$ are shown in~Fig.~\ref{K34_value}. We can find that the violation of LGI can approach the algebraic maximum of  $K_3^{max}\rightarrow3$ and $K_4^{max}\rightarrow4$ when  $\gamma\rightarrow J$. The algebraic bounds of $K_3$ and  $K_4$ in classical, Hermitian and non-Hermitian conditions can be summarized in Table~\ref{table1}.
\begin{table}
\caption{The algebraic bounds of LGI parameter $K_3$ and  $K_4$ in classical, Hermitian and non-Hermitian systems. It is noted that, for non-Hermitian case, the maximum and minimum values ​​of $K_4$  may not be obtained simultaneously with the certain observable.}
\begin{ruledtabular}
		\begin{tabular}{ccddd}
			Conditions &$K_3^{min}$&
			\multicolumn{1}{c}{$K_3^{max}$}&
			\multicolumn{1}{c}{$K_4^{min}$}&
			\multicolumn{1}{c}{$K_4^{max}$}\\
			\hline
			Classical system &-3&1& -2&  2 \\
			Hermitian system&-3&1.5& -2\sqrt{2}&  2\sqrt{2} \\
		
			Non-Hermitian system&-3& 3 \footnote{The value can be achieved when $\gamma\rightarrow J$. \label{a}} & -4 \textsuperscript{\ref{a}} & 4 \textsuperscript{\ref{a}}
		\end{tabular}
	\end{ruledtabular}
\label{table1}
\end{table}

\begin{figure}
	\centering
	\includegraphics[width=1.0\linewidth]{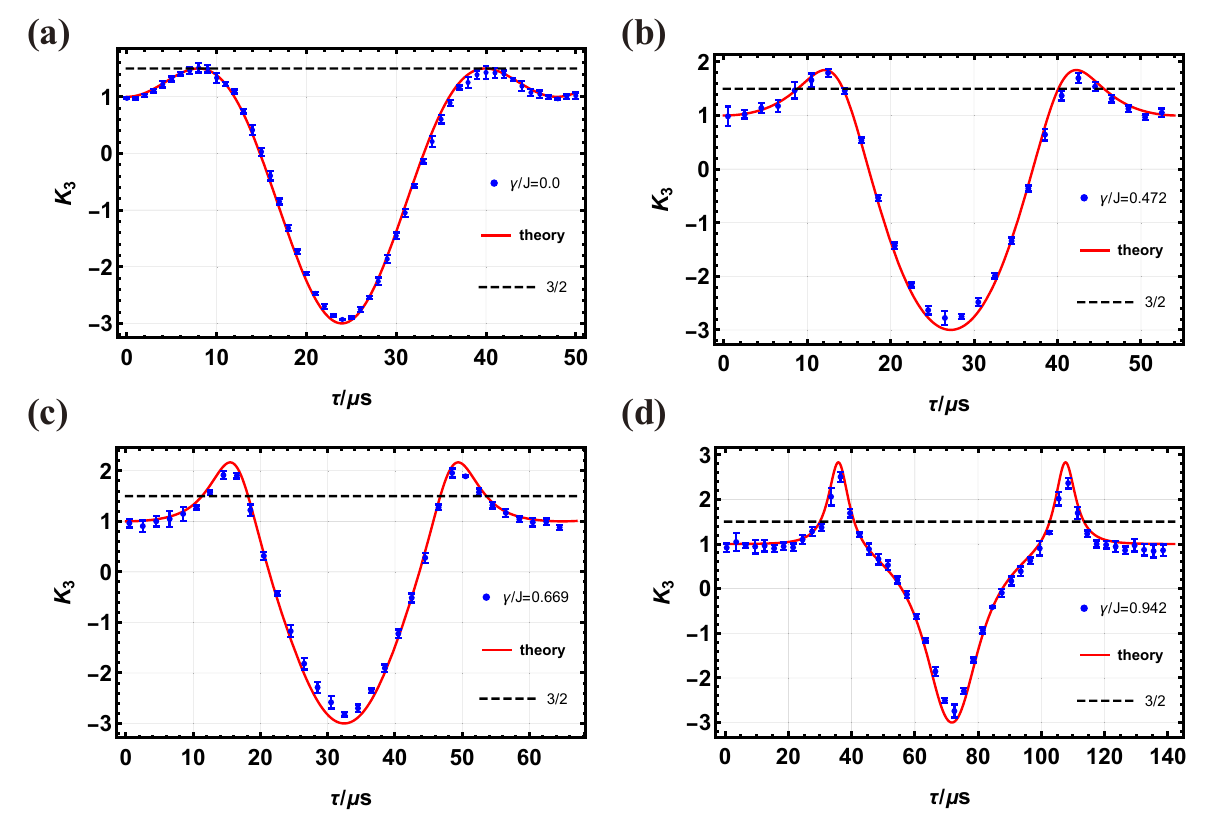}
	\caption{The experimental results of the LGI parameter $K_3$. (a) The measurement result of $ K_{3} $ under Hermitian condition ($\gamma/J=0$). (b-d) The measurement result of $ K_{3} $ under non-Hermitian condition, the ratios of $ \gamma/J $ are 0.472 (b) , 0.669 (c), and 0.942 (d), respectively.The red solid line in (a) (b-d) are the theoretical results calculated by Eq.~\ref{k3eq1} (Eq.~\ref{k3eq}), the blue circles are the experimental measured results, and the dashed line is the upper bound (3/2) of $K_3$ in the quantum system. The error bars of experimental results are estimated by the standard deviation (1$\sigma$) of multiple rounds of experiments. In order to obtain the discriminative population information of qubits, the number of repetitions of each set of experiments is 500. The dissipation rate in (a-d) corresponds to four dashed lines in Fig.~\ref{K34_value}(a).}
	\label{k3_results}
\end{figure}	
Fig.~\ref{k3_results} displays the experimental results of LG parameter $K_3$ when the system is undergoing the Hermitian dynamics and $\mathcal{PT}$-symmetric dynamics. In the Hermitian case, we adjust the applied duration of the microwave to get the correlation function (See Appendix~\ref{appC}) and demonstrate the $K_3$ curve for a qubit under Hermitian dynamics, as depicted in Fig.~\ref{k3_results}(a). This result agrees with the theoretical calculation of the red solid line. In addition, in order to get the $K_3$ curve in non-Hermitian condition, we need to simultaneously apply the microwave and dissipation beam to conduct non-unitary evolution. The $K_3$ curves of different dissipation intensities are shown in Fig.~\ref{k3_results}(b-d). These results show that in the non-Hermitian system, the upper bound of $K_3$ is violated beyond the limit allowed in the Hermitian system. The interpretation of the violations can be attributed to the non-Hermiticity, which changes the evolution speed of the quantum state~\cite{varma2022essential}. The phenomenon of the uneven evolution have been first observed in our previous experimental work~\cite{lu2022realizing}.
\begin{figure}
	\centering
	\includegraphics[width=1.0\linewidth]{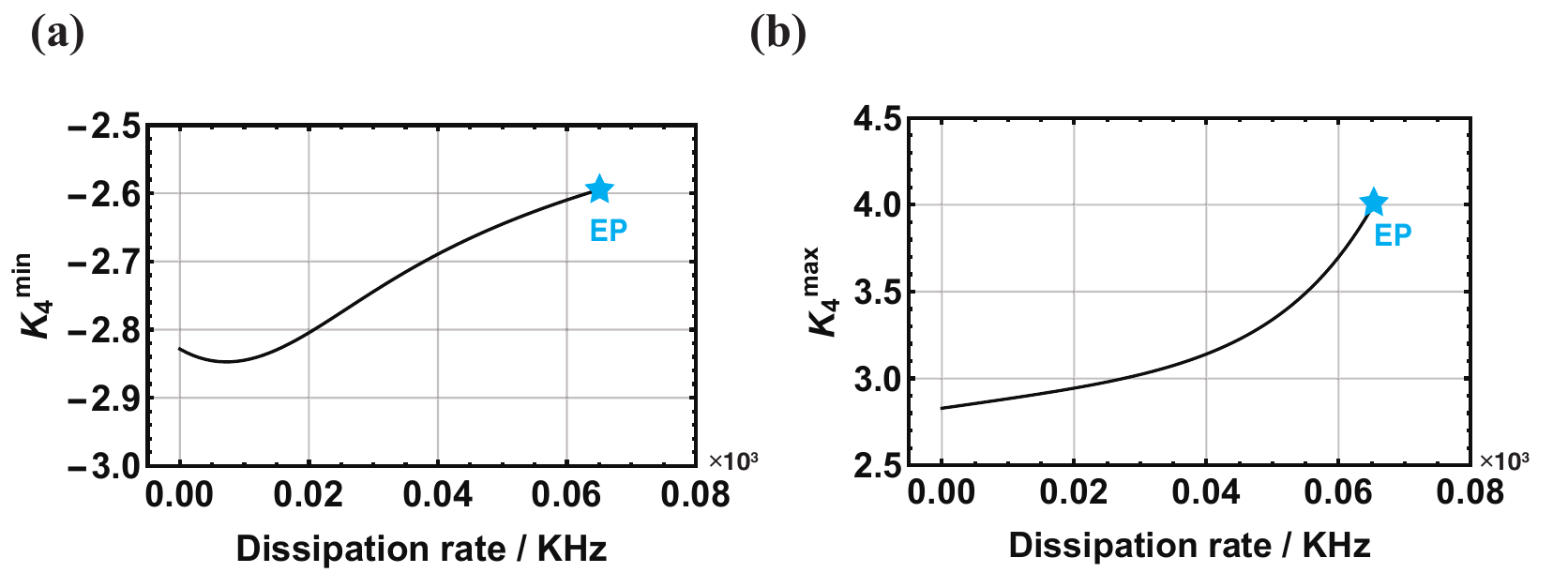}
	\caption{The numerical results of the higher-order LGI  $K_4$. (a) The lower bound and (b) the upper bound of  $K_4$ vary with the dissipation rate. The coupling strength $J$ in (a) and (b) are 2$\pi\times$10.4 KHz. The blue pentagram is the exceptional point (EP).}
	\label{k4_results1}
\end{figure}	

To further demonstrate the different behavior on the higher-order LGI ($K_4$) in non-Hermitian quantum system, we present the numerical and experimental results of $K_4$, as shown in Fig.~\ref{k4_results1} and~\ref{k4_results}. The protocol of $K_4$  involves measuring four two-time temporal correlations (See Appendix~\ref{appC}). Fig.~\ref{k4_results}(a) displays the experimental results under Hermitian dynamics, while Fig.~\ref{k4_results}(b-d) illustrate the enhanced violations of the upper bound for $K_4$ as the dissipation intensity increases. The lower bound of $K_4$ shows an upward trend with the increase of dissipation (Fig.~\ref{k4_results1}(a)), which is different from the case where the lower bound of $K_3$ remains constant. When the system approaches the exceptional point, the shift of the boundaries of  $K_4$ strongly depend on the target state, as shown in Fig.~\ref{optimize} in Appendix~\ref{appB}. It is worth noting that, when the dissipation increases, the upper and lower bounds may violate only one of the limits or both of the limits imposed by Hermitian condition, which
depends on the choice of the observable. This finding indicates that higher-order LGIs possess a distinct nature and offer certain advantages in some application scenarios~\cite{avis2010leggett,wilde2012addressing}.
\begin{figure}
	\centering
	\includegraphics[width=1.0\linewidth]{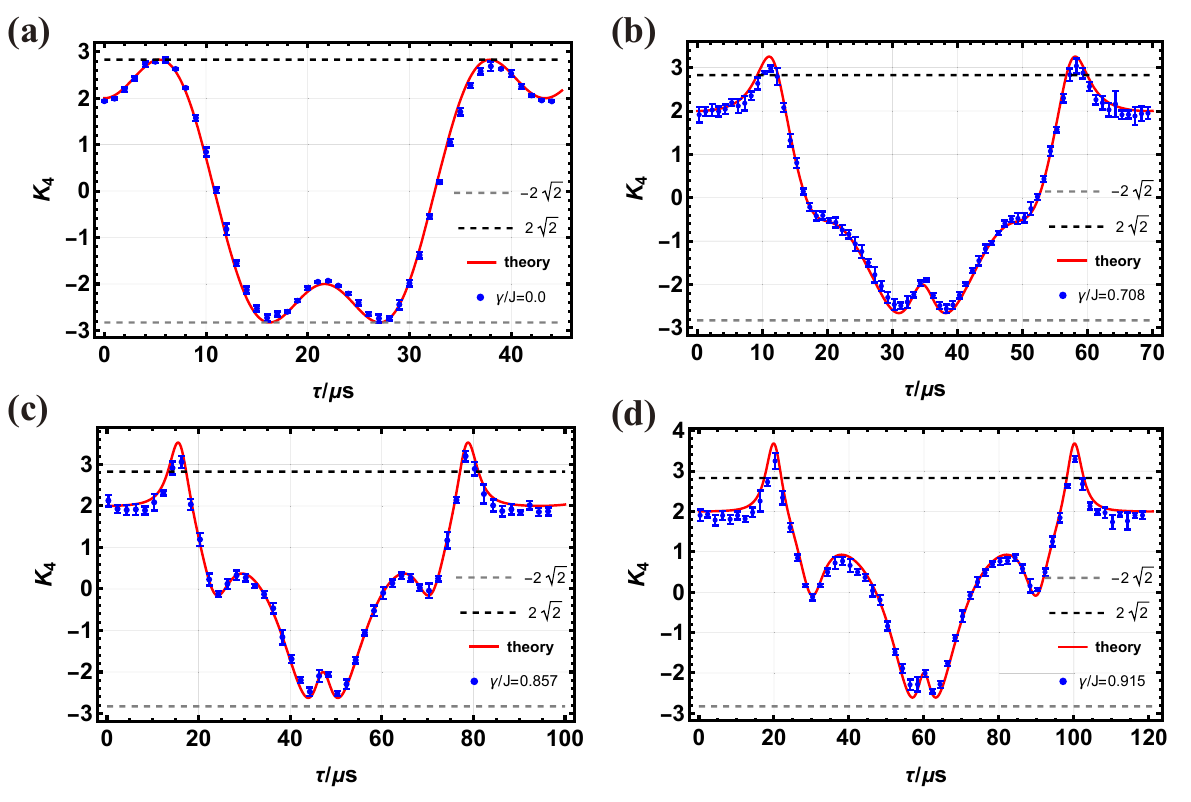}
	\caption{The experimental results of the LGI parameter $K_4$. (a) The measurement result of $ K_{4} $ under Hermitian condition ($\gamma/J=0$).  (b-d) The measurement result of $ K_{4} $ under non-Hermitian condition, the ratios of $\gamma/J $ are 0.708 (b), 0.857 (c), and 0.915 (d), respectively. The red solid line in (a) (b-d) are the theoretical results calculated by Eq.~\ref{k3eq1} (Eq.~\ref{k4eq}), the blue circles are the experimental measured results, and the black (gray) dashed line is the upper (lower) bound 2$\sqrt{2}$ (-2$\sqrt{2}$) of $K_4$ in the quantum system. The error bars of experimental results are estimated by the standard deviation (1$\sigma$) of multiple rounds of experiments. In order to obtain the discriminative population information of qubits, the number of repetitions of each set of experiments is 500. The dissipation rate in (a-d) corresponds to four dashed lines in Fig.~\ref{K34_value}(b).}
	\label{k4_results}
\end{figure}	

\section{Discussion and Conclusion}
In our previous work~\cite{lu2022realizing}, we have experimentally verified the faster than Hermitian dynamics using a trapped-ion qubit with a constructed $\mathcal{PT}$-symmetric Hamiltonian. The evolution time $|1\rangle\rightarrow |0\rangle$ gets shorter with increasing dissipation strength in the $\mathcal{PT}$-symmetric phase regime. The fast dynamics can be attributed to the nonlinear interaction in the dissipative two-level system, which shift the upper bound of LGIs~\cite{varma2021temporal}. Here, we start from the master equation to analyze this nonlinear interaction.
The qubit evolution three-level system can be described by ($\hbar= 1$)
\begin{equation}
	\frac{{d\rho }}{{dt}} =  - i\left[ {{H_{C}},\rho } \right] + \left( {{L_1}\rho {L_1}^\dag  - \frac{1}{2}\left\{ {{L_1}^\dag {L_1},\rho } \right\}} \right)
	\label{master}
\end{equation}
where $\rho \left( t \right)$ is a $3\times3$ density matrix, ${H_{C}} = J\left( {\left|  1\right\rangle \left\langle 0 \right| + \left| 0 \right\rangle\left\langle 1 \right|} \right)$ is a coupling Hamiltonian. ${L_ 1 }=\sqrt{4\gamma} \left| a \right\rangle \left\langle  1 \right|$ is the dissipation operator which accounts for the population probability decay from level $|1\rangle$  to $|a\rangle$ ($|a\rangle=^2S_{1/2}|F=1,m_F=\pm1\rangle$), where $4\gamma$ is the effective dissipation rate from $|1\rangle$ to $|a\rangle$.

\begin{figure}
	\centering
	\includegraphics[width=0.9\linewidth]{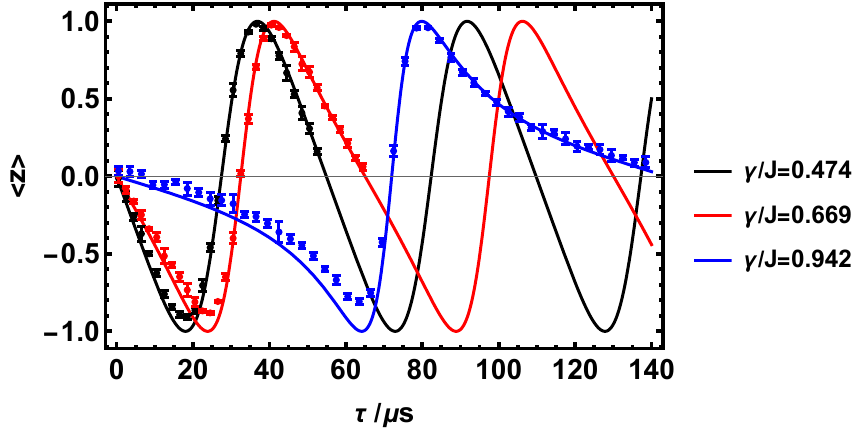}
	\caption{The experimental results of state evolution dynamic. The three solid lines are theoretical results, and different colors represent different dissipation strengths.}
	\label{speed}
\end{figure}

Rearranging the Eq.~\ref{master}, we obtain the $d\rho/dt=-i\left ( H_{eff}\rho-\rho H^\dagger_{eff}  \right) +4\gamma L_1\rho L^\dagger_1$, where the $H_{eff}=H_C-2i\gamma L^\dagger _1L _1$.  Because the microwave drive and the dissipation laser beam only act on the qubit, the dynamics of $|a\rangle$ is decoupled from qubit states, therefore we focus on the qubit subspace of the whole system, where the qubit retains its coherence and its dynamics is governed by an effective $\mathcal{PT}$-symmetric Hamiltonian, so we ignore the quantum jump term $4\gamma L_1\rho L^\dagger_1$ and obtain the following
\begin{equation}
	\frac{d\rho}{dt} =\begin{pmatrix}
		i J (\rho_{01}- \rho_{10}) & iJ(\rho_{00}-\rho_{11})-2\gamma\rho_{01} \\
		-iJ(\rho_{00}-\rho_{11})-2\gamma\rho_{10} & -iJ(\rho_{01}-\rho_{10})-4\gamma\rho_{11}
		\label{master1}
	\end{pmatrix}
\end{equation}

when we normalize the density matrix $\rho$, the above equation can be repalced by $d\rho/dt =-iJ\left [ \sigma_x ,\rho \right] +\gamma \left \{ \sigma_x,\rho \right \} -2\gamma \left \langle z \right \rangle\rho$. The Bloch component $\left \langle z \right \rangle=Tr[\sigma_z\rho]$ can be used to characterize nonlinear interaction~\cite{brody2012mixed,wu2023maximizing}. This interaction leads to a nonuniform evolution dynamic, as shown in Fig.~\ref{speed}. We can find that as the intensity of dissipation increases, the speed of state evolution is extremely inhomogeneous. The evolution time from $|+\rangle$ to $|0\rangle$ is long, but the evolution time from $|0\rangle$ to $|-\rangle$  is short. This shows a chiral property due to the unequal path in the non-Hermitian Bloch sphere~\cite{lu2022realizing}.

In summary, we have experimentally demonstrated the enhanced violations of the upper bound of both $K_3$ and $K_4$ of LGIs in a $\mathcal{PT}$-symmetric trapped-ion qubit. We also explored how the lower bounds for $K_3$ and $K_4$ depend on the order of LGIs. Notably, the lower bound for $K_3$ exhibits a constant behavior, while the lower bound for $K_4$ could shift an upward trend as dissipation increased. Furthermore, the competition between coherent dynamics and dissipative processes in open quantum systems makes it essential to establish the relationship between the LGI and the quantum speed limit~\cite{varma2022essential}. This relationship is crucial to understanding the enhanced violation of LGIs. In the future, it would be beneficial to investigate the LGI test in quantum many-body systems with dissipation, as this would open doors to studying non-Markovian dynamics~\cite{souza2013experimental} and quantum phase transitions~\cite{gomez2016quantum} in an open system.

\section*{acknowledgements}
This work is supported by the Key-Area Research and Development Program of Guangdong Province under Grant No.2019B030330001, the National Natural Science Foundation of China under Grant No.11974434 and No.12074439, Natural Science Foundation of Guangdong Province under Grant 2020A1515011159, Science and Technology Program of Guangzhou, China 202102080380, the Shenzhen Science and Technology Program under Grant No.2021Szvup172,  JCYJ20220818102003006, and Guangdong Science and Technology Project under Grant No.20220505020011. Le Luo acknowledges the support from Guangdong Province Youth Talent Program under Grant No.2017GC010656.

\appendix
\section{Experimental setup}\label{appA}
In order to confine a single $^{171}$Yb$^+$ ion in a linear Blade trap, we apply radio frequency (RF) signal and direct current (DC) voltages to two RF electrodes (RF1 and RF2) and two DC electrodes (DC1-5 and DC6-10), respectively. Each DC electrode is divided into five segments to compensate the excess micromotion and separate the frequency of radical mode~\cite{liu2021minimization}.The extra DC voltages (DC11 and DC12) applied on the RF electrodes are used to move ions to the RF null position.  We define trap axis as $\textbf{X}$ axis, and the other two axes perpendicular to the trap axis as $\textbf{Y}$ and $\textbf{Z}$ axis, respectively. The B field is along vertical $\textbf{Z}$ axis, which not only shifts the degeneracy of the three magnetic levels, but also prevents the ion from getting pumped into a coherent dark state~\cite{berkeland2002destabilization}. The generation of the 12.643 GHz microwave signal by mixing a standard RF source with either a direct digital synthesizer (DDS) or an arbitrary waveform generator (AWG). The resulting signal is then amplified to drive the transition between the hyperfine clock states ($|0\rangle$ and $|1\rangle$) by microwave horn.
\begin{figure}
	\centering
	\includegraphics[width=1.0\linewidth]{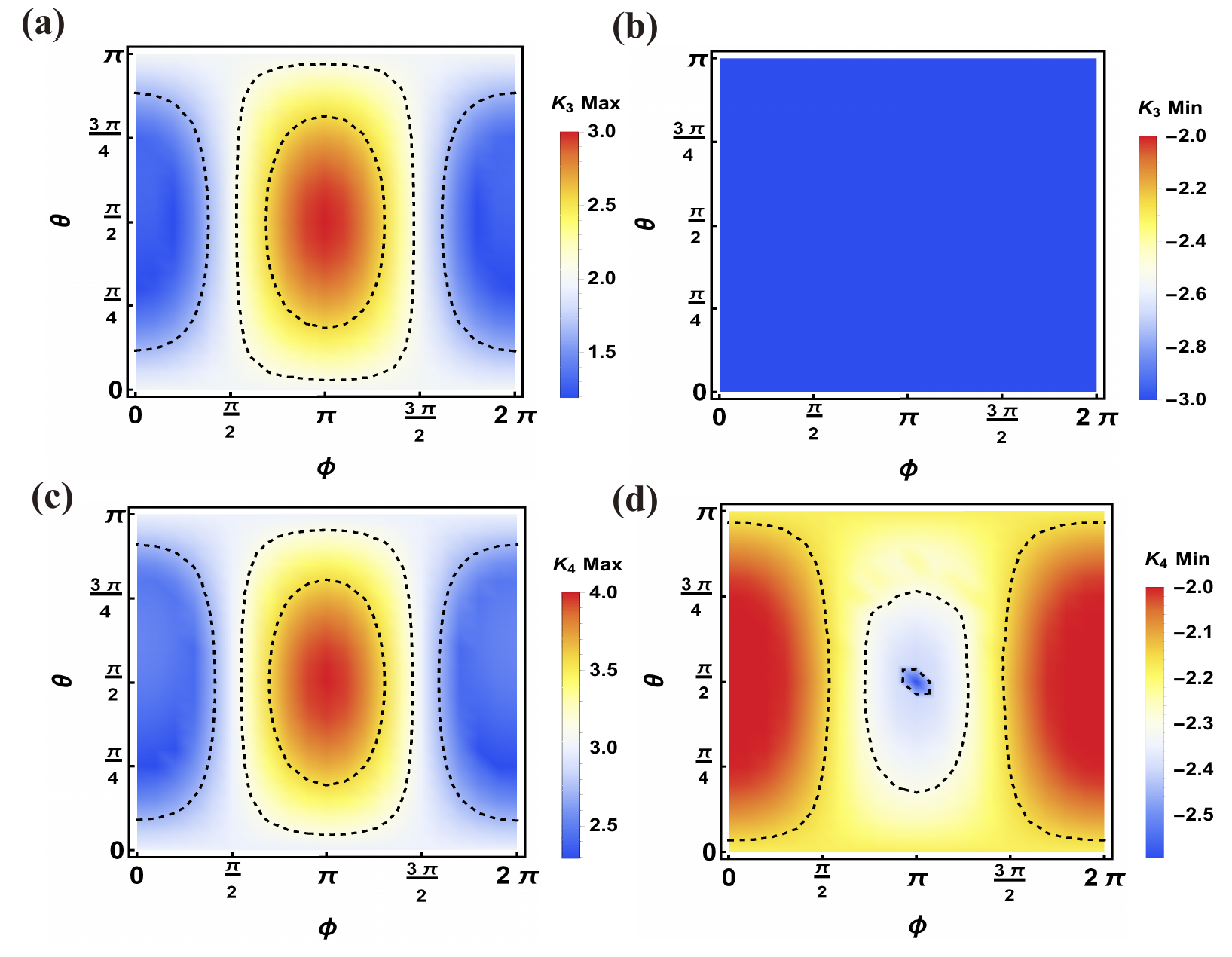}
	\caption{The numerical results for the maximum and minimum value of $K_3$ (a-b) and $K_4$ (c-d) by optimizing the parameter $\theta$ and $\phi$, where the ratio of the coupling strength and dissipation strength approaches the unity ($\gamma/J \rightarrow 1$). Here, the observable operator is $Q=\sigma_y$.}
	\label{optimize}
\end{figure}	
The timing sequences of the initialization (including cooling and optical pumping), dissipation beam, and detection beam are controlled by acoustic-optic modulators (AOMs) with RF switches, which receive TTL (Transistor Transistor Logic) signals from the ARTIQ (Advanced Real-Time Infrastructure for Quantum physics) device, as shown in Fig.~\ref{experiment_setup}(b). The synchronization of the microwave and the dissipation laser is precisely controlled by simultaneously triggering and passing through the same length of RF cables. The intensity of the dissipation beam can be adjusted by varying the RF power applied on the AOM. However, the population of the qubit will decay rapidly for a large $\gamma/J$ due to spin-dependent loss, resulting in a very small state population that is difficult to detect experimentally. To tackle this issue, a piecewise strategy is adopted, where the entire evolution time $T$ is divided into $N$ segments. In the $n$th ($1\le n\le N$) segment, the qubit is prepared to the state predicted by theory, which allows to evolve for $t_{n-1}=(n-1)T/N$ time. It then continues to evolve for $t=T/N$ under the non-Hermitian Hamiltonian $H_{eff}(t_n)$. This scheme allows the entire process to be mapped out effectively.

\begin{figure}
	\centering
	\includegraphics[width=1.0\linewidth]{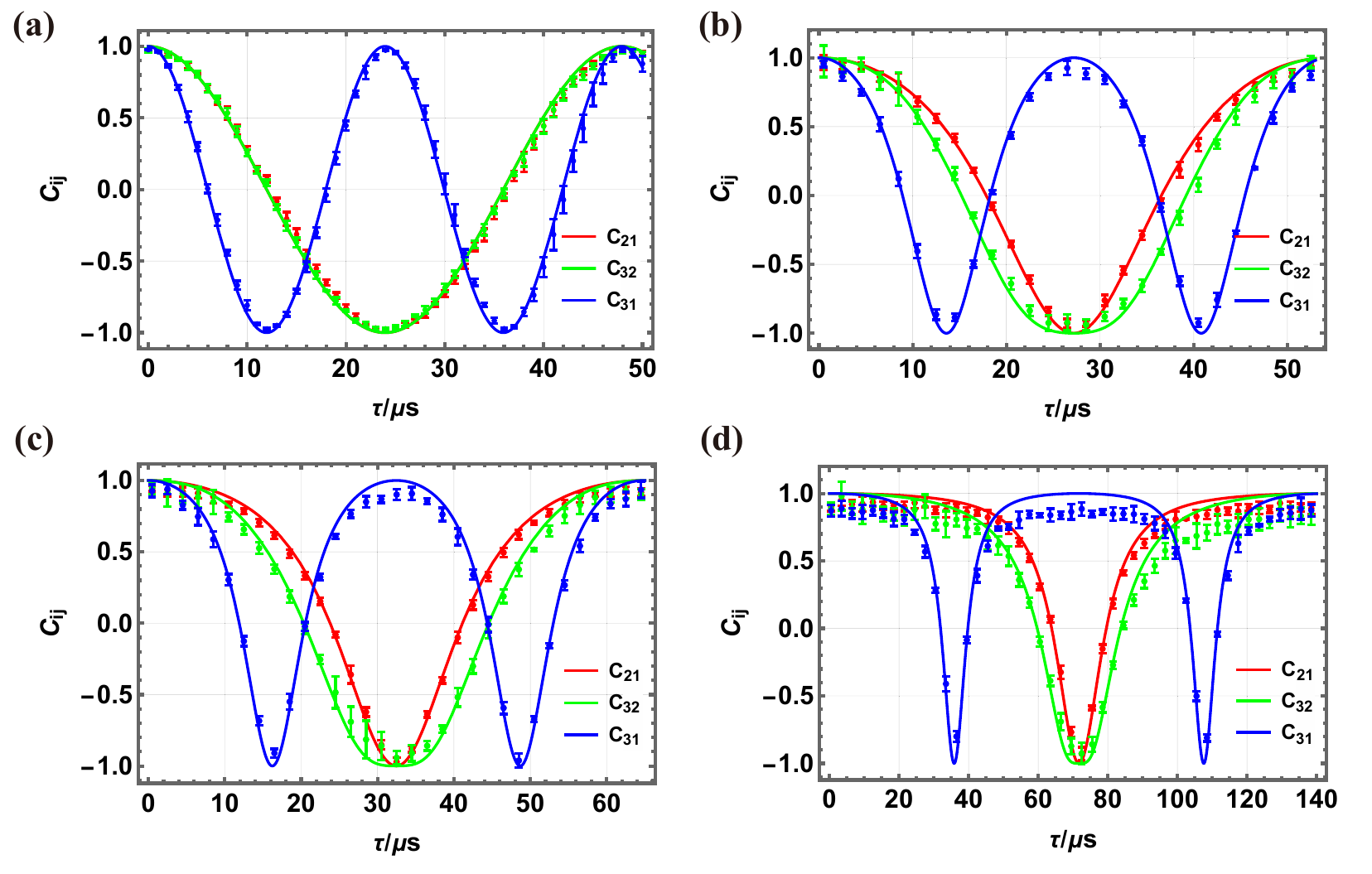}
	\caption{The experimental results of correlation functions $C_{ij}$ in LG parameter $K_3$.  The dissipation intensity $\gamma$ in four pictures are (a) 0, (b) 0.472 $J$, (c) 0.669 $J$ and (d) 0.942 $J$, respectively, where the coupling strength $J$ is 0.065 MHz.}
	\label{k3cij}
\end{figure}	

\begin{figure}
	\centering
	\includegraphics[width=1.0\linewidth]{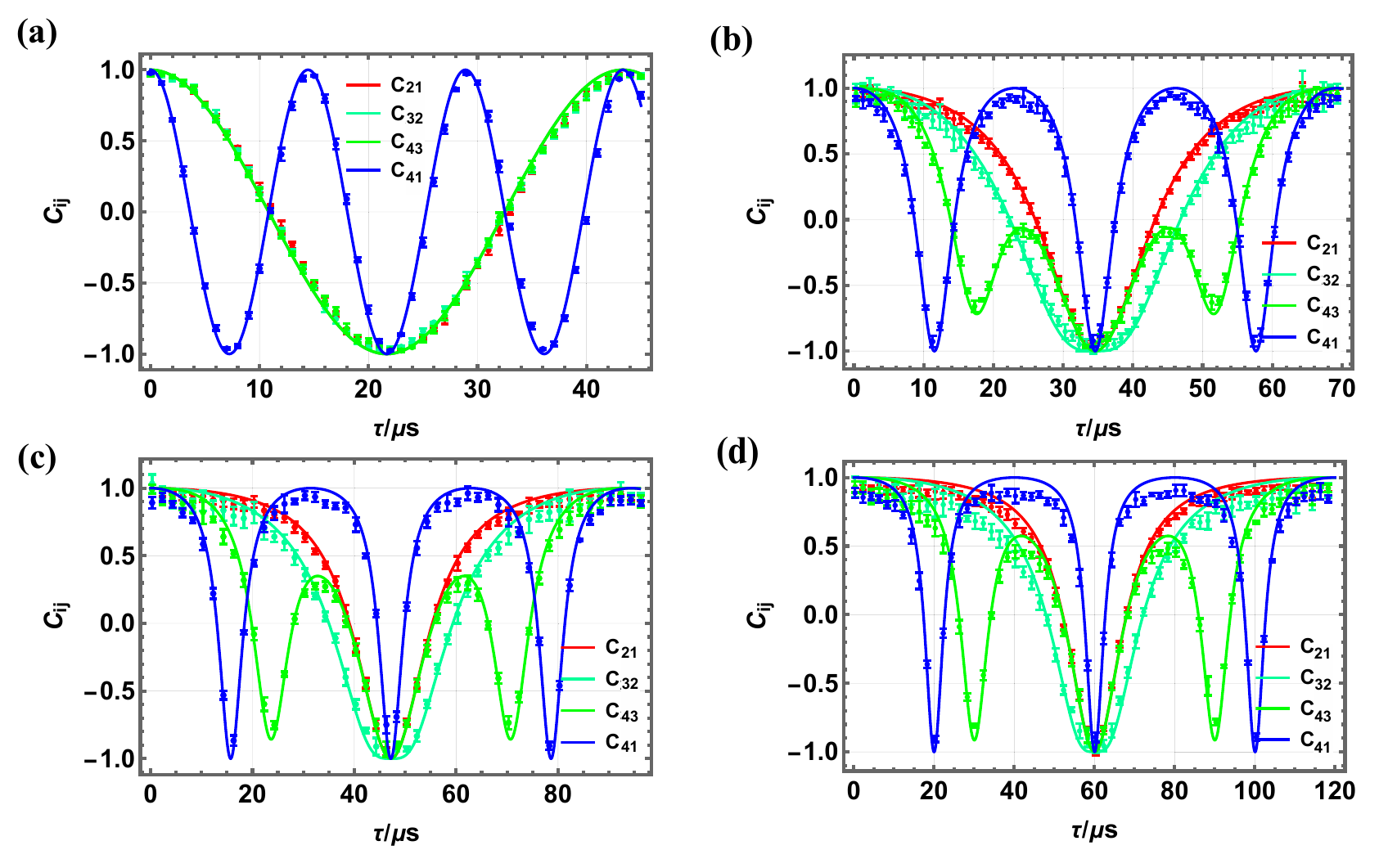}
	\caption{The experimental results of correlation functions $C_{ij}$ in LG parameter $K_4$.  The dissipation intensity $\gamma$ in four pictures are (a) 0, (b) 0.708 $J$, (c) 0.857 $J$ and (d) 0.915 $J$, respectively, where the coupling strength $J$ is 0.065 MHz.}
	\label{k4cij}
\end{figure}	

\section{The optimization of the target state }\label{appB}
The selection of the target state is crucial for carrying out an experimental test of LGIs in a dissipative quantum system~\cite{varma2022essential}. In our study, we begin by preparing the qubit in the initial state $|1\rangle$, which serves as the $|\psi_i\rangle$ state in Fig.~\ref{K34_scheme}. Next, we flip the $|\psi_i\rangle$ state to the target state $|\psi_t\rangle$ through the evolution operator $U(\theta ,\phi )$ in Eq.~\ref{U}.
\begin{equation}
U(\theta ,\phi )=\begin{pmatrix}
 \cos \frac{\theta}{2}  &-i\sin \frac{\theta}{2}e^{-i\phi } \\
 -i\sin \frac{\theta}{2}e^{i\phi } &\cos \frac{\theta}{2}
\end{pmatrix},
\label{U}
\end{equation}	
where the $U(\theta ,\phi )$ is obtained from the interaction Hamiltonian $H_I=J (\sigma_+e^{i\phi }+\sigma_-e^{-i\phi } )(\hbar=1)$ in the rotating frame. Here $\theta=2J t$ and $\phi$ are the rotating angle and rotating axis on the Bloch sphere. Thus, the target state $|\psi_t\rangle$ can be described as:
\begin{equation}
|\psi_t\rangle=U(\theta ,\phi )|\psi_i\rangle=\begin{pmatrix}
-ie^{-i\phi }\sin \frac{\theta}{2} \\\cos\frac{\theta}{2}
\end{pmatrix}.
\end{equation}	

The goal is to improve the correlation function with respect to the observable $Q=\sigma_y$ by optimizing the parameters $\theta$ and $\phi$. To achieve this, we have computed the maximum and minimum values of $K_3$ and $K_4$, and presented them in Fig.~\ref{optimize}. We find that the parameters of $\theta=\pi/2$ and $\phi=\pi$ are optimal solution of  $K_3$ and $K_4$ when $\gamma\rightarrow J$. Interestingly, the eigenstate $|+\rangle$ of the observable operator $Q$ is exactly the optimal target state.

\section{The calculations of two-time temporal correlations}\label{appC}
The two-time temporal correlations $C_{ij}$ of observable $\sigma_y$  can be written as:

\begin{widetext}
\begin{equation}
	\begin{aligned}
		&C_{21}=\frac{\gamma +J \cos \left(2 \tau \sqrt{J^2-\gamma ^2}\right)}{J+\gamma  \cos \left(2 \tau \sqrt{J^2-\gamma ^2}\right)}
		\\&C_{31}=\frac{\gamma +J \cos \left(4 \tau \sqrt{J^2-\gamma ^2}\right)}{J+\gamma  \cos \left(4 \tau \sqrt{J^2-\gamma ^2}\right)}
			\\&C_{41}=\frac{\gamma +J \cos \left(6 \tau \sqrt{J^2-\gamma ^2}\right)}{J+\gamma  \cos \left(6 \tau \sqrt{J^2-\gamma ^2}\right)}
		\\&C_{32}=\frac{J\gamma^2+J(J^2+J\gamma-\gamma^2)\cos(2\tau \sqrt{J^2-\gamma ^2} )}{(J-\gamma\cos(2\tau \sqrt{J^2-\gamma ^2} ))(J+\gamma\cos(2\tau \sqrt{J^2-\gamma ^2} ))^2}
		\\&-\frac{\gamma\cos(2\tau \sqrt{J^2-\gamma ^2} )(-J^2+J\gamma+\gamma^2+J^2\cos(2\tau \sqrt{J^2-\gamma ^2} ))}{(J-\gamma\cos(2\tau \sqrt{J^2-\gamma ^2} ))(J+\gamma\cos(2\tau \sqrt{J^2-\gamma ^2} ))^2}
		\\&C_{43}=\frac{2 \left(J^2-\gamma ^2\right) (\gamma +2 J) \cos \left(2 \tau \sqrt{J^2-\gamma ^2}\right)}{4 \left(J^2-\gamma ^2 \cos ^2\left(2 \tau \sqrt{J^2-\gamma ^2}\right)\right) \left(\gamma  \cos \left(4 \tau \sqrt{J^2-\gamma ^2}\right)+J\right)}
		\\&+\frac{\gamma  \left(2 J (J-\gamma ) \cos \left(4 \tau \sqrt{J^2-\gamma ^2}\right)+2 \left(J^2-\gamma ^2\right) \cos \left(6 \tau \sqrt{J^2-\gamma ^2}\right)-J \left(-2 \gamma +J \cos \left(8 \tau \sqrt{J^2-\gamma ^2}\right)+J\right)\right)}{4 \left(J^2-\gamma ^2 \cos ^2\left(2 \tau\sqrt{J^2-\gamma ^2}\right)\right) \left(\gamma  \cos \left(4 \tau \sqrt{J^2-\gamma ^2}\right)+J\right)}
	\end{aligned}
\end{equation}	
\end{widetext}

The experimentally measured temporal correlations $C_{ij}$ are depicted in Fig.~\ref{k3cij} and Fig.~\ref{k4cij}. The experimental results agree well with the corresponding theoretical predictions.

\end{document}